# Accuracy Threshold for Quantum Computation

Emanuel Knill[1]*, Raymond Laflamme[2] †, Wojciech Zurek[2]‡
[1] CIC-3, MS B265, [2] T-6, MS B288
Los Alamos National Laboratory, NM 87545, USA
and
Institute for Theoretical Physics,
University of California, Santa Barbara, CA 93106-4030, USA.

May 1996, revised October 1996

**Abstract**

We have previously [11] shown that for quantum memories and quantum communication, a state can be transmitted over arbitrary distances with error $\epsilon$ provided each gate has error at most $c\epsilon$. We discuss a similar concatenation technique which can be used with fault tolerant networks to achieve any desired accuracy when computing with classical initial states, provided a minimum gate accuracy can be achieved. The technique works under realistic assumptions on operational errors. These assumptions are more general than the stochastic error heuristic used in other work. Methods are proposed to account for leakage errors, a problem not previously recognized.

## 1 Introduction

Three recent events are promising to make extensive quantum computations as practical as classical computations. The first is the discovery by Shor [13], Steane [15] and Calderbank et al. [4, 3] of quantum error-correcting codes which can be used to maintain a quantum state for long periods of time or for long distances, assuming that the requisite recovery (or error-correction) operations can be implemented arbitrarily well. The second is the application of concatenated coding techniques by Knill and Laflamme [11] to quantum communication and memories. They demonstrate that a state can be maintained for an arbitrarily long time or distance at error $\epsilon$ provided each operation is implemented with error at most $c\epsilon$ for some constant $c$. The third is a proposal by Shor [14] to make a computation fault tolerant provided operations can be implemented with polylogarithmically bounded error. The results of [11] hold under realistic independence assumptions on the errors, while those of [14] were established for the stochastic error heuristic (i.e. random independent bit flip and sign flip errors).

*email: knill@lanl.gov
†laflamme@lanl.gov
‡whz@lanl.gov

Here we show that the techniques of concatenation [11], transversal implementation of encoded operations on linear codes [14, 16] and recovery based on purified states [14] can be combined to obtain arbitrary accuracy when computing with classical initial states, provided each operation can be implemented with better than a threshold accuracy. Thus the following statement holds under assumptions to be discussed:

**Theorem 1.1** *There exists a constant $\delta$ such that for every $\epsilon > 0$, a quantum algorithm using perfect operations (see below) can be converted to an equivalent quantum algorithm with imperfect operations, each with error at most $\delta$, such that the final error is at most $\epsilon$. The overhead of the converted algorithm is polylogarithmic in $1/\epsilon$ and the number of computation steps.*

In [14], $\delta$ depends polylogarithmically on $\epsilon$ and the complexity of the algorithm. For the purposes of this report, a quantum algorithm is a sequence of two qubit quantum operations starting with a classical initial state and ending in a measurement in the classical basis. This does not include algorithms which operate on unknown states, provided (for example) by a quantum oracle, unless those states are already in a suitably encoded form. Otherwise an initial encoding error cannot be avoided.

It is worth analyzing the assumptions under which Theorem 1.1 holds. A detailed treatment of the relevant assumptions is given in [10]. A summary follows. If the error in each operation is completely arbitrary, it is clearly not possible to prove any general fault tolerance results. In [14], the errors are assumed to occur stochastically and independently after correct operations. The errors are also assumed to belong to the error basis consisting of Pauli operators. This is called the stochastic independent error heuristic for the bit flip/sign flip error basis (the stochastic error heuristic, for short). It should be noted that although this heuristic is believed to yield reasonably good predictions of error behavior, it is usually not a good approximation to the actual error operators ( an exception is decoherence with memory operations only). In [11], the errors are assumed to be locally and sequentially independent. Here we argue that it suffices to assume that if the error behavior is expanded in an error basis, the total strength of errors involving any given $k$ error locations (both spatial and temporal) can be bounded by $\delta^k$, for some sufficiently small $\delta$. This assumption generalizes both the stochastic error heuristic and the local and sequential independence assumptions.

It is important to realize that the above assumptions on errors do not include leakage errors such as those which occur when the two-level approximation breaks down in an ion-trap. Such errors introduce new terms into the error operator expansion and must be dealt with by different methods. In particular, the methods described in [14, 11] are not robust against leakage without modification. We know of three methods to reduce or eliminate the effect of leakage. The first is to use a "stop leak" gate to return leaked amplitudes to the qubit. If such gates satisfy the accuracy requirements, this idea can be effective for non-concatenated methods of fault tolerant computation. The second is to explicitly represent the states to which amplitude can leak and encode them. The known fault tolerant methods can be applied, but the increase in dimensions of the primitive system increases the stringency of the accuracy requirements. The third involves the exploitation of detection/correction hierarchies together with a method for detecting leaks to control such errors, provided a threshold accuracy is satisfied. This method can be used for concatenated fault tolerant computation. Detection/correction hierarchies will be briefly discussed in Section 5.

Another assumption required for the formal analysis of fault tolerant computation concerns the effect of measurement on a qubit. It can be shown that the outcome of the measurement can be modeled by an inaccurate identity operation followed by a perfect measurement operation. However, it is useful to assume that even after an imperfect measurement, the qubit can no longer interact coherently with the remainder of the computational state. This effect can be achieved by allowing the qubit to decohere completely before reusing it.

In Section 2 we summarize the properties of simple concatenated coding without recoveries using purified states. It is shown that concatenated coding can be used to simplify the construction of a quantum memory by reducing the accuracy requirements and removing the need for specially designed recovery operations. This works provided that the top level encoding can reliably correct a constant error rate. With independent errors, the latter can be achieved by concatenated codes, even though in general they do not have good minimum distance. In Section 3 we show how encoded operations distribute into a concatenated code; and that if the punctured code construction is used, measurements and state preparation can be performed reliably in both the standard and the dual basis. We give a simplified implementation of primitive gates provided that the basic code satisfies that its weights are divisible by 8 and the dual has minimum distance 4. We also show how to use an apparently non-transversal implementation of a basic gate fault tolerantly. This trick can be used to avoid state preparation and measurement when implementing unitary primitives and allows the use of smaller codes. In Section 4 we view the techniques of transversally encoded operations and recovery using purified states as an error reduction technique that can be applied recursively by full concatenation. This reduces the accuracy requirements from those established by Shor [14] to yield the threshold result. Intuitive arguments are provided and an overview of the formal reasoning is given. In Section 5 we summarize the five known techniques supporting fault tolerant computing, including an alternative method based on error detection/correction circuits motivated by [5]. This method does not rely on purification for recovery. Details will appear in the full paper.

A threshold result for quantum computation with a version of the stochastic error heuristic has also been shown by Aharonov and Ben-Or [1]. They generalize the work of Gács [6] which uses concatenation for fault tolerant classical computation to deal with errors occuring during all, including the classical, operations. The give properties of encoded operations and recovery procedures that ensure the success of concatenation. In [1] classical operations are cast directly in terms of unitary operations, to avoid making any assumptions on measurements. As a result they do not take advantage of the known accuracy of classical operations. Concatenation for fault tolerant quantum computation with constant stochastic error is also used independently by Kitaev [8]. Kitaev relies on bounded width and depth syndrome checks to ensure fault tolerance of the recovery operators.

## 2 Concatenated codes

We review the concatenated quantum codes from [11]. The basic idea is to hierarchically code each qubit and interlace the procedure with recoveries in such a way that errors during recovery do not propagate as they would using simple repeated recovery operations. The technique recursively applies the coding procedures, where at the lowest level waiting periods (for communication or storage) or encoded operations (for fault tolerant computing) are applied between recovery operations. At the higher level, the recursive procedure is applied to each qubit between recovery operations. Although this allows only very limited application of encoded gates (due to indirect error propagation), this does simplify the quantum communication and memory problem. To be able to do extensive encoded computations requires more sophisticated recovery methods which will be covered in subsequent sections.

To perform concatenated coding, we choose an error-correcting code for a qubit (or other system) of length $l$ (i.e. using $l$ qubits) and a repetition factor $r$. The repetition factor is taken as large as reasonable subject to constraints given below. The length of the code is largely irrelevant, what matters is how much error per qubit can be tolerated at a low overall error after recovery.

The lowest level procedure ($\text{CCP}_r(1)$, for Concatenated Coding Procedure of depth 1 with $r$ repetitions) consists of simple iterated recoveries between an encoding and a decoding operation. That is, $\text{CCP}_r(1)$ begins with one qubit, encodes it using the error-correcting code to $l$ qubits, applies a recovery procedure to the code $r - 1$ times and finally decodes it back to a single qubit[1]. In between recovery operations, we can either just wait for

---

[1] The repetition factor is $r$ because the final decoding operation is a special form of the

a certain time interval, transmit each qubit over some distance, or apply a few suitably encoded operations involving other encoded qubits.

The higher level procedures $\text{CCP}_r(h)$ are defined recursively, using a procedure like $\text{CCP}_r(1)$, but with the next lower level applied to each qubit between recoveries. That is $\text{CCP}_r(h+1)$ starts with one qubit, encodes it using the code, applies $\text{CCP}_r(h)$ to each of the qubits of the code and recovers the code $r-1$ times, applies $\text{CCP}_r(h)$ to each qubit again and finally decodes the state to one qubit.

The error-correcting properties of $\text{CCP}_r(h)$ are discussed in [11]. In summary, suppose that the following holds: If each qubit of the code is subjected to independent interactions of error amplitude $e_d$, then the total error after recovery or decoding is at most $e_c$. Suppose also that $r+1 \leq e_d/e_c$. Then, if the error introduced in each qubit in $\text{CCP}_r(h)$ between recovery operations is independent and bounded by $e_d$, the final error of the result is at most $(r+1)e_c$. An important property of this technique is that no assumptions are made on the code used. The error propagation assumptions are all strictly worst case—no classical approximation is used which assumes that errors occur stochastically (see [11]). Also, any sufficiently high fidelity code can be used, not just $e$-error-correcting codes.

The total number of intervals between recoveries at the lowest level of the $\text{CCP}_r(h)$ is $r^h$, the total number of parallel recovery operators is $O(r^h)$, and the maximum number of qubits required is $l^h$, where $l$ is the length of the code. Thus, if the number of time intervals for which the state needs to be maintained is $n$, then the total overhead in qubits is $O(n^{1+c})$, with $1+c = \log_r l$. An explicit relationship between the error amplitude per operation and the overall error amplitude of the state can be found in [11].

Using concatenated codes as suggested above is very helpful for applications where each qubit to be preserved can be treated independently. However, to apply it to a quantum memory used during quantum computations requires more explanation. Preserving a qubit to within bounded error is only useful if that error does not propagate to the full state of the computation. Shor [14] has shown how to accomplish this by using recovery operations based on purified states. Thus, if the state of the computation is already encoded in a code which can tolerate error $\epsilon$ per qubit, then one can recovery operation, so in effect, $r$ recovery operations are used.

store each of the qubits of the encoded state using a concatenated code which meets this requirement without losing recoverability and fault tolerance of the encoded state. The code of the computation must be able to tolerate a constant error rate per qubit. Luckily, concatenated codes achieve this in principle, provided the errors are effectively independent. Each concatenation reduces the uncorrectable error of the lower level by a power related to the error-correction capacity of the basic code. The primary problem with the concatenated coding procedure as described above is that the qubit overhead can be relatively large. This problem is eliminated by the more sophisticated concatenation procedures to be described in Section 4.

## 3  Operating on states encoded in concatenated codes

A critical issue in operating directly on encoded states is to ensure that errors introduced during operations are corrected by recovery operators. In current proposals, this is achieved by ensuring that the encoded operations are transversal to the codes in the layout pattern of the qubits. Transversality can be defined as follows: Suppose that we wish to operate on $m$ qubits and that they are encoded using a code of length $l$. The complete set of qubits can be placed into an array of dimensions $l \times m$, where each column consists of the qubits supporting a coded qubit[2]. If the code can correct any $e$ errors, then the entire array is a code which can correct any type of error spanning at most $e$ qubits in each column. Somewhat loosely speaking, if we can implement encoded operations in such a way that non-local errors extend for more than $e$ qubits in a column at sufficiently low amplitudes, then recovery operations applied to each column can reliably restore the intended state.

A pictorial way to define transversality is to look at the dependencies of the encoded operations determined by the graph of the target qubits of each primitive operation that is applied. That is, connect qubits $x$ and $y$ if one of the operations targets both $x$ and $y$. The connected components of the resulting graph represent the potential dependencies between qubits, and it is these components that we must attempt to limit in their extent in each column. If the code corrects one error, the constraint requires that each primitive operation targets only qubits within a row (in some permutation of the elements of each column). We therefore refer to this as the *transversality* constraint. It is possible to relax these constraints if suitable codes are used, maybe even non-block codes[3].

---

[2]These columns are referred to as "qubytes" by Zurek and Laflamme [16]. They can be considered as the "qublocks" of a quantum block code.

Consider the possibility of weakening the transversality constraint while still ensuring that an encoded operation does not introduce uncorrectable errors at high amplitudes. If each operation is followed immediately by a recovery step, then the errors preceding the application of the encoded operation are essentially restricted to those types which can be introduced by the previous recovery operation. The encoded operation has the effect of spreading both pre-existing errors and (if there are dependencies) those introduced during the operation. The spreading is strongly restricted by the pattern of operations that are applied, and by the ordering of the operations. For example, consider applying controlled-nots to pairs of qubits corresponding to the edges in a path. If the desired effect requires applying the controlled-nots sequentially along the path, a bit flip error introduced by the first operation can be propagated to an equal amplitude error spread across the whole path. Other errors remain localized. If the desired effect can be achieved by applying the controlled-nots in two parallel steps of independent operations, then error propagation is restricted to at most three qubits at a time (for the dominant terms), and these triple errors are strongly constrained in terms of where they can occur. If this is known beforehand, the decoding procedure can be adapted to correctly decode such errors, even if not all triples can be corrected. We will give a concrete example of how knowledge of error propagation can be exploited.

The set of *encodable primitive operations* determines how computations are actually implemented. We use a simpler set of primitive operations than Shor [14] in order to avoid introducing complicated states to implement the Toffoli gates.

$$A = \begin{pmatrix} 1 & 1 \\ 1 & -1 \end{pmatrix},$$

---
[3]Using coding techniques other than non-block codes might be substantially more efficient, but may result in more difficult to implement encoded operations.

$$B = \begin{pmatrix} 1 & 0 \\ 0 & i \end{pmatrix},$$

$$C = \begin{pmatrix} 1 & 0 \\ 0 & -i \end{pmatrix},$$

$$D = \begin{pmatrix} 1 & 0 & 0 & 0 \\ 0 & 1 & 0 & 0 \\ 0 & 0 & 1 & 0 \\ 0 & 0 & 0 & i \end{pmatrix},$$

$$E = \begin{pmatrix} 1 & 0 & 0 & 0 \\ 0 & 1 & 0 & 0 \\ 0 & 0 & 1 & 0 \\ 0 & 0 & 0 & -i \end{pmatrix},$$

$$N = \begin{pmatrix} 1 & 0 & 0 & 0 \\ 0 & 1 & 0 & 0 \\ 0 & 0 & 0 & 1 \\ 0 & 0 & 1 & 0 \end{pmatrix}.$$

The operations $D$ and $E$ are controlled phase shifts by $i$ and $-i$, respectively, using the standard lexicographic labeling of the classical states of two qubits. $N$ is a controlled-not. According to Shor [14], it suffices to use $A$ and $D$ to obtain a sufficiently dense set of operations for quantum computation, because a Toffoli gate can be simulated. We added the others so that the set of operations used in the encoding is identical to that being encoded. This simplifies the analysis of the implementation of the operations on concatenated codes. To see that our set of operations is sufficiently dense we observe that $D^2$ is a controlled sign flip, $A^\dagger D^2 A$ (with $A$ acting on the second qubit) is a controlled-not, the two-controlled sign flip can be implemented using two controlled-nots, two $E$'s and one $D$ using the well-known trick of Barenco et al. [2], and the Toffoli gate can be obtained from the two-controlled sign flip the same way as the controlled-not can be obtained from the controlled sign flip and $A$. The claim then follows from Shor's statement in [14].

An important issue is how well the desired operations can be approximated by primitive encodable ones. In order for these methods to work, the ideal computation without errors in the primitives must be sufficiently close to the desired one. According to a naive argument, the approximation by the encodable primitive operations of the constructs of a computation must be within at least $1/n$ (or at least $1/\sqrt{n}$, optimistically), where $n$ is the length of the computation. It is therefore important either to have a rich enough set of encodable primitives, or to be able to efficiently approximate the desired set of computational constructs[4].

We do not know of any codes which easily permit transversal encoding of the complete set. However, as we will see, $B$, $C$, $D$, $E$ and $N$ as well as state preparation of $|+\rangle = |0\rangle + |1\rangle$ (we leave out normalization factors) and measurement in the $|+\rangle, |-\rangle = |0\rangle - |1\rangle$ basis can be accomplished transversally. This suffices for computing $A$ in an effectively transveral way by using a simple version of the trick used to implement Toffoli gates in [14]. Suppose that a qubit is in state $|\psi\rangle$, and we wish to obtain $A|\psi\rangle$. We first introduce a second qubit in state $|0\rangle + |1\rangle$. Applying a controlled-not controlled by the second qubit yields the state $|\psi\rangle|0\rangle + \begin{pmatrix} 0 & 1 \\ 1 & 0 \end{pmatrix} |\psi\rangle|1\rangle$. Apply $B$ to the second qubit and change basis to obtain

$$|\psi\rangle(|+\rangle+|-\rangle) + \begin{pmatrix} 0 & i \\ i & 0 \end{pmatrix} |\psi\rangle(|+\rangle-|-\rangle) = \begin{pmatrix} 1 & i \\ i & 1 \end{pmatrix} |\psi\rangle|+\rangle + \begin{pmatrix} 1 & -i \\ -i & 1 \end{pmatrix} |\psi\rangle|-\rangle.$$

Measuring in the $|+\rangle, |-\rangle$ basis thus yields either $X = \begin{pmatrix} 1 & i \\ i & 1 \end{pmatrix}$ or $Y = \begin{pmatrix} 1 & -i \\ -i & 1 \end{pmatrix}$, and we know which one. We have $-i \begin{pmatrix} 0 & 1 \\ 1 & 0 \end{pmatrix} X = Y$, and $BYB = A$.

Codes which permit transversal implementation of operations $B$, $C$, $D$, $E$ and $N$ exist. Let $C$ be a (classical) code with weights divisible by 8 and whose dual $D$ has minimum distance at least 4. The Reed-Muller codes $\mathrm{RM}(r, m)$ have this property for $m \geq 3r + 1$ and $r \geq 1$ (with dual having minimum distance 4) and $r \geq 2$ (with dual having minimum distance 8) [12]. For example, $\mathrm{RM}(1, 4)$, is a code of length 16 which works. Let $C'$ be the punctured code, with $C'_0$ the even subcode. If $D'$ is the punctured dual code, then $C'^{\perp}_0 = D'$ and $C'^{\perp}$ is the subcode $D'_0$ of $D'$ derived from the vectors

---
[4]R. Solovay and A. Yao (unpublished) have apparently shown that any dense generating set of operations permits polylogarithmic approximation of unitary two-qubit operations. Knill and Laflamme (unpublished) have independently found a constructive method based on Kitaev's [7] ideas. The latter method works for the set of primitives given above.

with a 0 in the deleted bit. We encode $|0\rangle$ and $|1\rangle$ as suggested in [15, 14]:

$$|0_L\rangle = \sum_{c \in C'_0} |c\rangle, \quad |1_L\rangle = \sum_{c \in C'_1} |c\rangle,$$

where $C'_1$ denotes the odd subset of $C'$. Note that if we apply the full Hadamard transform $H$ to the state we obtain

$$H|+_L\rangle = \sum_{d \in D'_0} |d\rangle, \quad H|-_L\rangle = \sum_{d \in D'_1} |d\rangle,$$

where $D'_1$ is the complement of $D'_0$ in $D'$. $H$ is of course obtained by applying $A$ to each qubit. Thus a measurement in the $|+_L\rangle, |-_L\rangle$ basis is obtained by Hadamard transforming each qubit and measuring in the classical basis. This is equivalent to measuring each qubit in its $|+\rangle, |-\rangle$ basis. A measurement is automatically fault resistant, provided that the inferred state is deduced by using a classical syndrome decoding method which matches the one used to correct the quantum errors in both bases.

Because $C$'s weights are divisible by 8, $C'$'s weights are either 0 or $7 \bmod (8)$, depending on whether the word is in $C'_0$. This immediately allows encoding $B$ by applying $C$ to each qubit and encoding $C$ by applying $B$ to each qubit. The not operation can also be effected by applying it to each qubit. If $D$ (or $E$ or $N$) is applied individually to each corresponding pair of qubits in two encoded states, then the operation $E$ (or $D$ or $N$, respectively) is applied to the encoded state. For $N$ this follows from the fact that $C'$ is closed under sums. To see that it works for $D$ and $E$, we show that if $x, y \in C'$, then the intersection of $x$ and $y$ is either $0 \bmod (4)$ or $3 \bmod (4)$, where the latter holds only if both are in $C'_1$. Let $|z|$ be the weight of a word $z$. Let $k$ be the overlap between $x$ and $y$. If $x, y \in C'_0$, then $|x + y| = |x| + |y| - 2k = 0 \bmod (8)$. Since $|x| = 0 \bmod (8)$ and $|y| = 0 \bmod (8)$, $k = 0 \bmod (4)$. If $x \in C'_0$ and $y \in C'_1$, then $x + y \in C'_1$, so $|x + y| = |x| + |y| - 2k = 7 \bmod (8)$ and $|y| = 7 \bmod (8)$ imply that $k = 0 \bmod (4)$. If $x, y \in C'_1$, then $x + y \in C'_0$. We have $|x + y| = |x| + |y| - 2k = 6 \bmod (8) - 2k = 0 \bmod (8)$, so $k = 3 \bmod (4)$.

The disadvantage of the method discussed above is that the smallest known code satisfying the required property has 15 qubits, while Shor's implementation of the Toffoli gate allows use of the 7 qubit Hamming code.

Luckily an alternative method exists which avoids the use of the measurement trick to implement one of the operations. The 7 qubit Hamming code is based on the 8 bit self-dual code $RM(1,3)$. This code satisfies that each code word's weight is divisible by 4. It follows that the 7 qubit quantum code permits transversal implementations of $A$, $B$, $C$ and $N$. $D$ can be implemented by applying 7 transversal stages in such a way that $D$ is applied to each pair of qubits with the first from the control qubyte and the second from the target qubyte. To ensure that errors do not propagate out of control, a partial recovery operation extracting and correcting only the bit flip errors is applied between each transversal stage. To see that this works requires two observations. The first is that since $D$ does not change the support of the code, a bit flip following transversal applications of $D$ can be detected and corrected even though the coding space has not been preserved. The second is that $D$ commutes with sign flip errors, so that such errors are not propagated by subsequent applications of $D$.

To apply the encoded operations to the concatenated codes, each encoded operation can be re-encoded at each level of the hierarchy. In effect, the operations are applied transversally to the leaf qubits of the code, with the desired effect. It is well worth considering the construction of the concatenated code directly in terms of the theory of linear codes. As an example, we describe the punctured code construction for one concatenation. Let $C_2$ consist of codewords obtained by first selecting a codeword of $C$, then replacing each bit other than the first one with either a codeword of $C'_0$ or $C'_1$, depending on whether the bit is 0 or 1. As can be seen, the construction used for quantum codes naturally appears in this context. The concatenated quantum code is obtained by puncturing $C_2$ at the bit that was not replaced. Note that the dual of $C_2$ is obtained by an identical construction: Take a codeword of $D$, then replace all but the bit to be punctured by a codeword of $D'_0$ or $D'_1$, depending on the value of the bit. As a consequence both $C$ and $D$ can be decoded very easily, simply by computing syndromes hierarchically.

## 3.1 State preparation and measurement

The complete set of operations required for quantum computation requires the ability to perform state preparation and measurement. Normally it suf-fices to be able to prepare the encoded classical state $|0_L\rangle$ and measure in the classical basis. For the implementation of $A$ by preparation and measurement, it is also necessary to prepare $|+_L\rangle$. For the punctured codes discussed here, measurement can be performed in the $|0_L\rangle, |1_L\rangle$ basis by measuring each qubit in its classical basis and assigning the outcome to $|0_L\rangle$ or $|1_L\rangle$ by computing the classical syndrome pattern and applying classical error correction. Similarly one can measure $|+_L\rangle$ versus $|-_L\rangle$ by first applying $A$ to each qubit to transform each qubit to the $|+\rangle, |-\rangle$ basis and then applying the same procedure using the classical dual code. This method is transversal, and the measurement is robust against errors.

State preparation can be performed by applying a fault tolerant method for extracting parity information. Shor's recovery method using purified states can be used for the purpose as discussed in [14]. Basically the idea is to detect and correct one more syndrome to force the contents of the qubit into the desired state if no error occurs, and to ensure that the dominant errors that are introduced behave independently. For example, in the case of the punctured codes, the final state can be forced into $|+\rangle$ with low residual error (depending on operational accuracy), because $|+\rangle$ is the only state which is supported in $C'$ in the classical basis and in $D'_0$ in the dual basis.

## 3.2 Fault tolerant recovery

We briefly review the fault tolerant recovery method of [14]. Although we are not planning on applying it to the concatenated code as a whole (because of the bottleneck induced by the need of extracting the correct syndrome with high amplitude), the details of the implementation for a single instance of a code has a substantial impact on the actual values that can be obtained for the threshold operational error. For our purposes, the most important instance of the method occurs when it is necessary to correct single errors to reduce the overall error from linear to quadratic.

Shor's fault tolerant recovery method involves reliably extracting syndrome information without introducing correlated errors by utilizing a specially prepared state. For each parity of $l$ bits that needs to be computed, the state $|e_l\rangle$ on $l$ qubits consisting of a uniform superposition of all even states needs to be prepared in such a way that it contains sign-flip errors with very small amplitude only. The parity is obtained by applying a transver-sal controlled-not operation from the coded state to $|e_l\rangle$ and is extracted by measuring $|e_l\rangle$ in the classical basis. The parities of the syndrome are computed several times by this method to gain enough confidence in having correct syndrome information. The even state is prepared by purifying "cat" states and applying $A$ to each qubit.

The simplest method to achieve quadratic error reduction requires preparing the cat state by initializing $l$ qubits in a linear arrangement in $|00\ldots\rangle$, applying $A$ to the first, then applying controlled nots in succession to adjacent qubits. This results in a superposition of the $|00\ldots\rangle$ and the $|11\ldots\rangle$ states with a potentially large number of errors. The "bad" errors for error propagation are the bit flips, which except for those which in the end only affect one qubit and those occurring in conjunction with other errors have been spread to the final qubit in the sequence. Such errors are reliably detected by comparing the first and last qubits using controlled-nots with a test qubit. This test has the useful property that the only correlated errors that are introduced to the cat state are sign-flips. If the test is successful, the cat state is then transformed by applying $A$ to each qubit before performing the parity computation.

This method for fault tolerant recovery is effective if not particularly efficient. A. Steane (unpublished) is investigating more efficient methods which require fewer interactions with the encoded state. In any case, the number of operations required for quadratic error reduction is $O(l^2)$, where $l$ is the length of the code. Quantum pseudocode with the details of the method will be provided in the full paper.

## 4  Fault tolerant recoveries as error reduction

Consider the application of transverally encoded operations and recovery with purified states to a specific code, which we require to correct at least one error. For our purposes, it is most useful to consider the Hilbert space of the supporting qubits of a code as a tensor product of an abstract two-state particle (with state $|0_L\rangle$ and $|1_L\rangle$), and the syndrome space (see [9]). Intuitively, the syndrome space acts as a protective barrier between errors introduced by the environment or the operations and the abstract particle, somewhat like a dam designed to hold back excess water in a downpour. If the syndrome space is in its initial state, errors affect only the syndrome, not the abstract particle. A useful property is that for the punctured coding construction, errors in the standard error basis which saturate the protection offered by the syndrome space can be expressed directly as one of the standard errors acting on the abstract particle. It is the abstract particle that we want to use for computations.

The state encoded using abstract particles defined by a code and the recovery method has two properties important to fault tolerant computing. The first is the fidelity of the recoverable state. Define the *recoverable state* to be the state induced on the abstract particles. The recoverability of a state compared to the intended state is the fidelity between the two. The *loss* is the associated error amplitude.

Consider the application of transversally encoded operations and recovery operations to a single instance of a one-error correcting code of length $l$. For analysis, it is convenient to hardwire the syndrome computations, for example by computing the whole set of parities three times, taking the first two or second two outcomes, if they agree, or doing nothing otherwise. If the amplitude of success of one set of syndrome computations is $p$, then the overall amplitude of failure is at most $O(p^2)$. Let us follow the procedure of taking incoming encoded qubits, applying a requested operation in the encoding as described in Section 3 followed by a full fault tolerant recovery operation. This involves $O(l^2)$ operations, except when the non-transversal implementation of $D$ is used, which requires $O(l^3)$ operations. If the incoming state has independent errors in the supporting qubits or cross qubyte pairs of qubits of amplitude $O(p)$, then its loss is $O(p^2)$, due to the error-correcting properties of the code. Applying the operations introduces independent single qubit errors at an amplitude of $O(p)$, the recovery operation fails with amplitude $O(p^2)$ and otherwise fixes the previous errors and introduces new independent error of $O(p)$ total amplitude. Thus the encoded operation and recovery acts with error bounded by $\alpha p^2$ on the abstract particle, where the constant $\alpha$ depends on the actual number of operations required to implement a single step.

We can view this as a method for utilizing a particle on which we can operate with error bounded by $p$ to obtain an abstract particle on which we are able to act with error bounded by $\alpha p^2$. The overhead of the method is bounded by $\beta l^2$ in operations and $\gamma l$ in qubits for some constants $\beta$ and $\gamma$. A back of the envelope calculation for the constants in the case of the 7 qubit codes gives $\beta \sim 60$. $\alpha$ can be calculated as the number of pairs of places an error can occur. This is roughly the square of the number of operations, in this case about $8 \cdot 10^6$. This is expected to improve substantially when fault tolerant computing methods are optimized, but suggests a threshold of around $10^{-6}$, about two orders of magnitude worse than that obtained in [11] for quantum channels. The next step is to amplify the accuracy by applying the same method to the abstract particles. A naive calculation shows that after $h$ concatenations, the new error per operation is bounded by $(\alpha p)^{2^h}$, while the overhead is bounded by $\beta^h l^{2h}$ and $\gamma^h l^h$, respectively. If the computation to be converted has $n$ steps, then it is necessary to satisfy $(\alpha p)^{2^h} \leq \epsilon/n$ ($\leq \epsilon/\sqrt{n}$ if we are optimistic, that is if we anticipate that errors behave according to the random walk model), which can be accomplished with $h \sim \log_2 \log_{1/(\alpha p)}(n/\epsilon)$, which for $p < \alpha$ results in a polylogarithmic overhead.

## 4.1 Overview of the formal analysis

The simplest situation for which the argument outlined above can be formalized correctly is under the stochastic error heuristic where only operations in the normalizer of the error group are used. This includes all of the basic operations except for $D$ and $E$. In that case the argument based on induction on the number of levels in the concatenation can be made to work, primarily because any uncorrected but stochastic sign and bit flips at the previous level can be expressed as stochastic sign and bit flips for the abstract particle of the next level. An important aspect of the analysis ensures that the syndrome spaces are maintained in states which can absorb sufficient numbers of additional errors without causing overflow into the abstract particles.

The basic technique for understanding error propagation is to consider specific occurrences of errors in the hardwired fault tolerant implementation of an operation with recovery. The error operators are placed on the connections between operations in the quantum network. To see whether the error is tolerated, one can conjugate each of these errors with the subsequent unitary operators to find that they either get eliminated by recovery operators or result in a correctable error pattern after the encoded operations. The conjugation operation replaces $EU$ with $UU^\dagger EU$, thus replacing the error $E$ before $U$ with the error $U^\dagger EU$ after $U$. A technical difficulty occurs if a $D$ or $E$ operator needs to be conjugated with a bit flip, as in that case $U^\dagger EU$ is a sum of elements of the error basis, rather than a stochastic combination. To prove that accuracy amplification still works requires a more careful analysis using weaker assumptions on how errors occur, or, alternatively, characterizing explicitly those minimal error patterns which cause the computation to fail. There are systematic methods for converting such a characterization into bounds on the error amplitude, assuming only that the total strength of correlated errors is exponentially small in the number of places in the network that are affected. In our experience, results which show a threshold error probability of $\delta$ under the stochastic error heuristic can usually be generalized to give a threshold error amplitude of $c\delta$ under the weakest independence assumptions. The value of $c$ depends on the nature of the error patterns which cause failure. There are artificial examples for an abstract version of the problem of relating stochastic arguments to amplitudes where $c$ is as small as $1/|\log(\delta)|$.

# 5 Summary of fault tolerant computation techniques

In general, the design goal of a fault tolerant network is to ensure that most errors that occur (stochastically or coherently) do not affect the outcome of the final measurement. We now know of five[5] techniques that can be used in combination to achieve this goal. These techniques are quantum coding, transversally encoded operations, concatenation, recovery with purified states and detection/correction. The first four have been shown to yield threshold results for quantum computation under general independence assumptions and without leakage errors. The results for quantum channels demonstrate that recovery with purified states can be avoided if no encoded operations or state preparations are required.

---

[5] The state of the art now includes several otheeer techniques. Aharonov and Ben-Or [1] define a general "bounded spread" property sufficient for fault tolerant recovery of coded states and Kitaev [8] uses codes with suitably bounded syndrome checks which have such a property. Another emergent principle involves using constraints on the physical errors as leverage to improve accuracy.

Detection/correction is a different method used on a small scale by Cirac, Pellizzari and Zoller [5] for correcting errors that occur in the phonon mode in ion traps. It can be used together with concatenation to amplify operational accuracy. The basic idea is to use an encoding of a qubit and associated recovery operation which permits detection of any one (say) error in such a way that if no error is detected, the state is known to have quadratically less error. This encoding is concatenated with one that permits restoration of the state if an error occurs in a known position. To get a quadratic accuracy gain, the top level code must be implemented so that recovery is only attempted if an error is detected in the lower level. In that case, recovery is performed directly without explicit concatenation of the operations or ancillas. Any single error detecting code can be used for both levels of the concatenation. However, care needs to be used in implementing the recovery operation to ensure that single errors anywhere in the circuit are detected. An important advantage of detection/correction is that leakage errors are much more naturally accounted for by this method, provided they can be detected with sufficiently small error on the physical qubits.

# 6 Conclusion

We have reported on work in progress to formally analyze the five fault tolerant computing techniques: quantum coding, transversally encoded operations, concatenation, recovery with purified states and detection/correction. So far we have established that strong threshold results hold for quantum computation and clearly delineated the assumptions which suffice for these results. We have offered a few methods for dealing with a potentially serious source of errors due to amplitude leakage from the physical qubits. A full report will appear.

# 7 Acknowledgments

Special thanks to Ben Schumacher and Richard Hughes. We have greatly benefited from interaction with the Quantum Computer group at Los Alamos

National Laboratory. We also thank Alexei Ashikhmin for his assistance with classical error correcting codes. This work was supported in parts under the auspices of the U.S. Department of Energy under Contract No. W-7405-ENG-36, and by the National Science Foundation under Grant No. PHY94-07194.